%% file: MS-cSLIC.tex
\newif\ifpreprint
\preprintfalse % Enable for double column reprint

\ifpreprint
\documentclass[journal=jpclcd,manuscript=letter]{achemso}
\else
\documentclass[journal=jpclcd,manuscript=letter,layout=twocolumn]{achemso}
\fi

\usepackage[T1]{fontenc} % Use modern font encodings

\usepackage{amsmath}
\usepackage{newtxtext,newtxmath}

\usepackage{pifont}
\usepackage{graphicx}
\usepackage{dcolumn}
\usepackage{braket}
\usepackage{multirow}
\usepackage{threeparttable}
\usepackage{xspace}
\usepackage{verbatim}
\usepackage[version=4]{mhchem} % Formula subscripts using \ce{}
\usepackage{comment}
\usepackage{color,soul}

\usepackage{mathtools}
\usepackage[dvipsnames]{xcolor}
\usepackage{xspace}
\usepackage{ifthen}

\usepackage{qcircuit}

\usepackage{graphicx,longtable,dcolumn,mhchem}
\usepackage{rotating,color}
\usepackage{lscape}
\usepackage{amsmath}
\usepackage{dsfont}
\usepackage{soul}
\usepackage{physics}

\usepackage[colorlinks = true,
            linkcolor = blue,
            urlcolor  = black,
            citecolor = blue,
            anchorcolor = black]{hyperref}
\definecolor{goodorange}{RGB}{225,125,0}
\definecolor{goodgreen}{RGB}{5,130,5}
\definecolor{goodred}{RGB}{220,50,25}
\definecolor{goodblue}{RGB}{30,144,255}

\newcommand{\note}[2]{
\ifthenelse{\equal{#1}{F}}{
\colorbox{goodorange}{\textcolor{white}{\footnotesize \fontfamily{phv}\selectfont #1}}
    \textcolor{goodorange}{{\footnotesize \fontfamily{phv}\selectfont #2}}\xspace
}{}
\ifthenelse{\equal{#1}{R}}{
\colorbox{goodred}{\textcolor{white}{\footnotesize \fontfamily{phv}\selectfont #1}}
    \textcolor{goodred}{{\footnotesize \fontfamily{phv}\selectfont #2}}\xspace
}{}
\ifthenelse{\equal{#1}{N}}{
\colorbox{goodgreen}{\textcolor{white}{\footnotesize \fontfamily{phv}\selectfont #1}}
    \textcolor{goodgreen}{{\footnotesize \fontfamily{phv}\selectfont #2}}\xspace
}{}
\ifthenelse{\equal{#1}{M}}{
\colorbox{goodblue}{\textcolor{white}{\footnotesize \fontfamily{phv}\selectfont #1}}
    \textcolor{goodblue}{{\footnotesize \fontfamily{phv}\selectfont #2}}\xspace
}{}
}

\usepackage{titlesec}

\usepackage[fontsize=11pt]{scrextend}
\titleformat{\section}
{\normalfont\sffamily\bfseries\color{Blue}}
{\thesection.}{0.25em}{\uppercase}

\titleformat{\subsection}[runin]
{\normalfont\sffamily\bfseries}
{\thesubsection}{0.25em}{}[.\;\;]

\titleformat{\suppinfo}
{\normalfont\sffamily\bfseries}
{\thesubsection}{0.25em}{}

\titlespacing*{\section}{0pt}{0.5\baselineskip}{0.01\baselineskip}
\titlespacing*{\subsection}{0pt}{0.125\baselineskip}{0.01\baselineskip}

\author{Mohamed Sabba}
	\affiliation{School of Chemistry and Chemical Engineering, University of Southampton, Southampton SO17 1BJ, UK}
\author{Christian Bengs}
	\affiliation{School of Chemistry and Chemical Engineering, University of Southampton, Southampton SO17 1BJ, UK}
\author{Urvashi D. Heramun}
	\affiliation{School of Chemistry and Chemical Engineering, University of Southampton, Southampton SO17 1BJ, UK}
\author{Malcolm H. Levitt}
   \email{mhl@soton.ac.uk}
	\affiliation{School of Chemistry and Chemical Engineering, University of Southampton, Southampton SO17 1BJ, UK}

\let\oldmaketitle\maketitle
\let\maketitle\relax
	\title{Error compensation without a time penalty:  robust  spin-lock-induced crossing in solution NMR}
\date{\today}

\begin{tocentry}
	\vspace{1 cm}
 \includegraphics[width=1\textwidth,trim={0 9cm 19cm 0},clip]{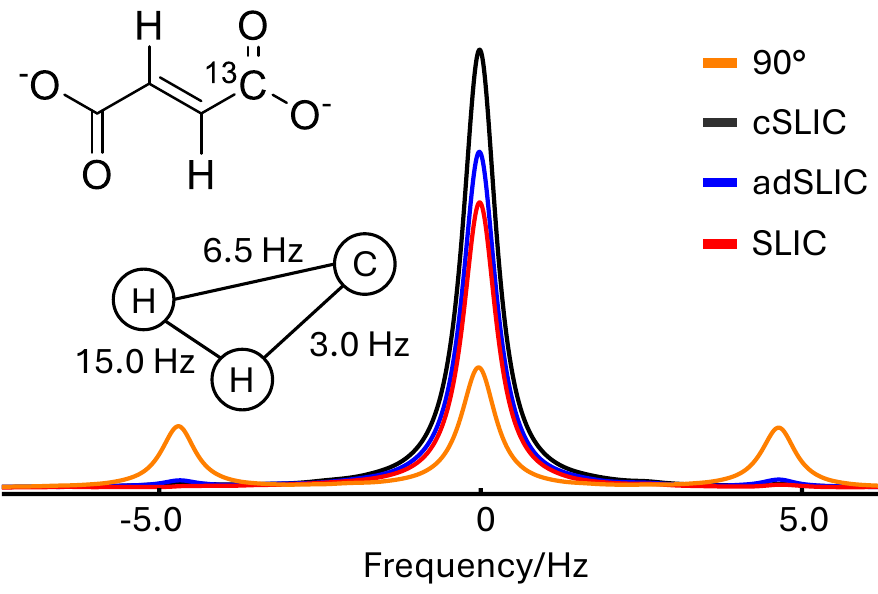}
\end{tocentry}

\begin{document}	

\ifpreprint
\else
\twocolumn[
\begin{@twocolumnfalse}
\fi
\oldmaketitle

%%%%%%%%%%%%%%%%
%%% ABSTRACT %%%
%%%%%%%%%%%%%%%%

\begin{abstract}
A modification of the widely-used spin-lock-induced crossing (SLIC) procedure is proposed for the solution nuclear magnetic resonance (NMR) of strongly coupled nuclear spin systems, including singlet NMR and parahydrogen-enhanced hyperpolarised NMR experiments. The compensated-SLIC (cSLIC) scheme uses a repetitive sequence where the repeated element employs two different radiofrequency field amplitudes. Effective compensation for deviations in the radiofrequency field amplitude is achieved without increasing the overall duration of the SLIC sequence. The advantageous properties of cSLIC are 
demonstrated by numerical simulations and by representative experiments. 
\end{abstract}

\ifpreprint
\else
\end{@twocolumnfalse}
]
\fi

\ifpreprint
\else
\small
\fi

\noindent

%%%%%%%%%%%%%%%%%%%%
%%% INTRODUCTION %%%
%%%%%%%%%%%%%%%%%%%%

\input{Symbols/main}
\input{Figures/ContourPlots}

\section{Introduction}

The spin-lock-induced crossing (SLIC) method was introduced by De~Vience \emph{et al.} in 2013 as a simple method for generating long-lived singlet order for near-equivalent spin pairs in nuclear magnetic resonance~\cite{devience_preparation_2013}.
SLIC involves the application of a resonant radiofrequency field with the same phase as transverse nuclear spin magnetisation. The amplitude of the resonant field is chosen so that the nutation frequency under the rf field matches the J-coupling between the members of the spin pair:
\begin{equation}
\label{eq:SLICcondition}
    \omega_\mathrm{nut}
    =\omega_J,
\end{equation}
where $\omega_J=2\pi J$. When applied to strongly coupled spin-1/2 pairs
%in the near-equivalence regime
$|\omega_\Delta|\ll|\omega_J|$, where $\omega_\Delta=2\pi\Delta$, and $\Delta$ is the chemical shift difference in units of Hz. This matching condition induces a level crossing in a suitable reference frame\cite{rodin_representation_2020} which leads to the coherent evolution of transverse magnetisation into nuclear singlet order (SO). Singlet order represents a net difference in populations of the singlet and triplet states of the spin-1/2 pair
\begin{equation}
\label{eq:SO_order}
Q_{\rm SO}=\vert S_{0}\rangle \langle S_{0}\vert-\frac{1}{3}\sum_{m}\vert T_{m}\rangle \langle T_{m}\vert
=-\tfrac{4}{3}\mathbf{I}_j\cdot\mathbf{I}_k
,
\end{equation}
where $\vert S_{0}\rangle$ and $\vert T_{m}\rangle$ are the singlet and triplet states formed by the two participating spins $I_j$ and $I_k$~\cite{levitt_singlet_2012,levitt_long_2019}. In suitable circumstances, singlet order is a ``long-lived state", protected against some common relaxation mechanisms, and which may exhibit a decay time constant exceeding 1 hour in favourable circumstances~\cite{stevanato_nuclear_2015}.
%\MHLnote{Someone (Chris?) is not using the zotero cSLIC library, so we have lost previous references!} 
%\CBnote{are the references correct? The MRI reference for example seem to refer to 31P NMR?} \SBnote{Yes. The 31P reference is one of the only ones using SLIC for spectral selection. I contacted Matt Rosen and he suggested these two recent SLIC MRI references} \SBnote{Contacted Kirill to see if he has any references to suggest, but we already had the papers cited}
%\newline\indent An important practical advantage of SLIC is that it relies only on a continuous-wave spin-locking field. 
%\MHLnote{we don't want to play that up too much since cSLIC is not so simple!}
%\CBnote{Makes sense, I've removed the last bit.}
%Consequently, 

SLIC has also been used for the selection of desirable NMR~\cite{devience_nmr_2021a} and MRI~\cite{boele_ultralow_2025,mcbride_scalable_2025} signals, the estimation of J-coupling differences on the order of a few mHz~\cite{devience_probing_2016}, for chemically informative J-spectroscopy in low magnetic fields~\cite{devience_homonuclear_2021,devience_homonuclear_2022,mandzhieva_zerofield_2025}, and for parahydrogen-enhanced NMR~\cite{theis_lightsabre_2014,pravdivtsev_spin_2014,eills_singlet_2017, knecht_efficient_2019, sheberstov_hyperpolarization_2021,dagys_robust_2024, boele_ultralow_2025}.
SLIC may also be used to manipulate multiple-quantum transitions in strongly coupled spin systems~\cite{sonnefeld_polychromatic_2022a,sonnefeld_longlived_2022,sheberstov_collective_2024,razanahoera_hyperpolarization_2024,wiame_longlived_2025}.

% However, most 
Many applications of SLIC are hampered by its high sensitivity to deviations in the rf field amplitude. This behaviour is illustrated by the contour plot in Fig.~\ref{fig:ContourPlots}(a), which shows the amplitude for the transformation of
%\MHLnote{amplitude for the transformation of?}
transverse nuclear magnetisation 
(described by the operator $I_x$) 
into nuclear singlet order (described by the operator $Q_{\rm SO}$). 
%\MHLnote{defining and using the transformation amplitude $<I_x\to-\tfrac{4}{3}\mathbf{I}_j\cdot\mathbf{I}_k>$ seems necessary here. Also, the form of the singlet order operator needs to be explained better. We probably need to first define the singlet and triplet states etc. }
The transformation amplitude is plotted as a function of two parameters. The horizontal axis shows the J-coupling normalised resonance offset
\begin{equation}
\Omega=\Omega_\mathrm{rf}-\tfrac{1}{2}\left(\Omega^0_j+\Omega^0_k\right),
\end{equation}
where $\Omega_\mathrm{rf}$ is the radio-frequency field frequency and $\Omega^0_j,\,\Omega^0_k$ are the chemically-shifted Larmor frequencies of the two spins. The vertical axis shows the fractional deviation of the rf field amplitude, expressed as a nutation frequency, $\omega_{\rm{nut}}$, from the nominal value $\omega_{\rm{nut}}^0$:
\begin{equation}
\label{eq:FractionalrfDeviation}
\epsilon_{\rm rf}=
\frac{\omega_{\rm{nut}}-\omega_{\rm{nut}}^0}{\omega_{\rm{nut}}^0}.
\end{equation}

The narrow crescent shape traced out by the contours in figure~\ref{fig:ContourPlots}(a) indicate that the performance of SLIC is very sensitive to the rf amplitude when applied at exact resonance, and also that the resonance offset and rf amplitude errors interact strongly. In general, these are undesirable characteristics. In particular, the high rf field sensitivity implies a degradation in performance whenever the rf field is inhomogeneous over the sample volume, as is often the case.
%\section{Radiofrequency field compensation}
%\newline\indent 

Several methods have been proposed for rendering SLIC less sensitive to amplitude errors.
%~\cite{theis_composite_2014}. 
One popular option is to extend the pulse sequence duration while varying the rf amplitude continuously so as to sweep through a range of matching conditions~\cite{theis_composite_2014,rodin_constantadiabaticity_2019}. This solution is often known as adiabatic-SLIC (adSLIC) even though the strict adiabaticity conditions on the spin dynamics are not always met~\cite{comparat_general_2009,rodin_constantadiabaticity_2019}. Simulations for one possible amplitude shape given by Theis and coworkers\cite{theis_composite_2014,claytor_making_2015} are shown in figure~\ref{fig:ContourPlots}(b), where the specific rf-amplitude modulation is given by:
\begin{align}\label{eq:AdPulseShape}
\omega_{\rm{nut}}(t) &=
\omega_{J} \left[1 - \Delta_{\rm{max}} \tan\left( x \xi \pi/2\right)/\tan\left(\xi \pi/2 \right) \right] , \nonumber \\
x &= 2t/T -1,
\end{align}
where the time variable $t$ runs from $0$ to the total duration $T$. The parameter $-1<\Delta_{\rm{max}} < 1$ defines the size and direction of the sweep, whereas $0<\xi<1$ is a shape parameter, with small values of $\xi$ converging to a linear sweep and large values leading to a more adiabatic transfer. 

As expected, sensitivity to the rf amplitude is reduced relative to unmodulated SLIC, albeit at the cost of a significantly increased duration, which can lead to increased losses through dissipation, although this is not taken into account in the calculations of figure~\ref{fig:ContourPlots}. In practice, this dissipation sets the ultimate limit on performance enhancement: while in principle, the robustness of adSLIC can be increased indefinitely by increasing the pulse duration, which is not possible in spin systems which relax rapidly. In the simulation of Figure~\ref{fig:ContourPlots}(b), a total adSLIC duration of $T=500$ ms was chosen, approximately seven times longer than the SLIC sequence, to ensure a fair comparison.

\input{Figures/cSLIC}
\input{Figures/FumaratePulseSequences}
\section{The \MakeLowercase{c}SLIC sequence}

We recently showed that SLIC may also be used for double-quantum excitation in systems of spin-1/2 pairs, and proposed a modified SLIC sequence, called compensated-SLIC or cSLIC, in order to improve its tolerance to errors~\cite{heramun_spinor_2026}. The cSLIC sequence was shown to provide effective compensation for rf amplitude deviations, without an increase in pulse sequence duration relative to the uncompensated SLIC method. Here we show that the cSLIC sequence is also significantly more robust than uncompensated SLIC in the context of singlet NMR, and in contexts such as parahydrogen-enhanced NMR. 
%\newline\indent 

The cSLIC sequence involves $n$ repetitions of the cyclic element $\mathcal{C}$ sketched in figure~\ref{fig:cSLIC}(a), which has the form:
\begin{equation}
\label{eq:cSLIC}
\mathcal{C} = 
(\alpha \pi)_x^\mathrm{weak} 
- (\alpha 2\pi)_{-x}^\mathrm{strong}-(\alpha\pi)_{x}^\mathrm{weak}.
\end{equation}
Compared to conventional SLIC, the cSLIC sequence employs two different amplitude levels: a weak rf field, %matching the SLIC condition
and a stronger rf field which provides a compensatory counter-rotation in the centre of each repeating element. The nutation frequency of the two weak outer pulses matches the SLIC condition:
\begin{equation}
  \omega^\mathrm{weak}_\mathrm{nut} =  \omega_J.
\end{equation}
By contrast, the strong central pulse is assumed to have a nutation frequency that is much larger than those of the outer pulses, $\omega^\mathrm{strong}_\mathrm{nut}>\omega^\mathrm{weak}_\mathrm{nut}$. The parameter $\tfrac12 <\alpha\lessapprox 1$ is given by
\begin{equation}
\label{eq:alpha factor}
\alpha = \frac{\omega^\mathrm{strong}_\mathrm{nut}}{
\omega^\mathrm{strong}_\mathrm{nut}
+\omega^\mathrm{weak}_\mathrm{nut}
}.
\end{equation}
In the limit of an infinitesimally short central pulse ($\omega^\mathrm{strong}_\mathrm{nut}\gg\omega^\mathrm{weak}_\mathrm{nut}$), the parameter $\alpha$ tends to 1, and the cSLIC sequence becomes
\begin{equation}
\label{eq:cSLIC-StrongPulseLimit}
\text{cSLIC($\alpha\to1$)} = 
(\pi)_x^\mathrm{weak} 
- (2\pi)_{-x}^\mathrm{strong}
-(\pi)_{x}^\mathrm{weak},
\end{equation}
where the central $2\pi$ pulse has negligible duration. In practice, this limit cannot be reached, due to hardware limitations, but it can be approached sufficiently closely.

The duration of each cSLIC element is given by $\tau_J=|J|^{-1}$ for all values of $\alpha$. In the absence of relaxation, 
the singlet-order generation for an ensemble of 2-spin-1/2 systems is maximised when $n$ is chosen as follows~\cite{heramun_spinor_2026}:
\begin{equation}
 n=\lfloor J/(\sqrt2\Delta)\rceil,
\end{equation}
where $\lfloor x \rceil$ rounds $x$ to its nearest integer.

The compensation principle can be illustrated in the limit $\alpha=1$, where the duration of the central compensating pulse is negligible compared to the element duration $\tau_J$. Assume that the sequence is applied on-resonance ($\Omega=0$). Under these ideal conditions, the two weak SLIC pulses combine to generate a $2\pi$ rotation, and the central pulse generates an equal and opposite $2\pi$ rotation. When the rf field amplitude of the weak pulses is larger than expected, $\omega^{\rm weak}_{\rm nut}=(1+\epsilon_{\rm rf})\omega_J$, the excess rotation induced by the misset SLIC field is  compensated by the equal and opposite excess rotation caused by the misset strong pulse -- and similarly when the rf field is weaker than nominal. This compensation mechanism assumes that the weak and strong rf fields experience the rf amplitude errors in the same proportion. This is the case when both fields are generated by the same radiofrequency coil, and when the deviations are caused by spatial variations in the rf field strength. This is the usual experimental situation. 

If the central pulse strength does not dominate the outer pulses ($\alpha<1$), the net rotations of the inner and outer pulses still compensate each other for any value of the overall rf amplitude, and the duration of each repeating element still matches one period $\tau_J$ of the J-coupling frequency. Hence, the basic physics of the level anticrossing~\cite{pravdivtsev_exploiting_2013,pravdivtsev_spin_2014,rodin_representation_2020} still applies, albeit with slightly reduced state-mixing coefficients. The simulation shown in figure~\ref{fig:ContourPlots}(c) confirms this analysis for the realistic case $\alpha=0.99$. 
%\MHLnote{?} \SBnote{corrected} 
The robustness of the singlet excitation amplitude with respect to rf amplitude deviations is greatly improved, at the expense of a somewhat reduced frequency bandwidth. The undesirable interaction between rf amplitude and resonance offset effects is strongly reduced. 
The improved error compensation of cSLIC is achieved with the same sequence duration as uncompensated SLIC.

\input{Figures/FumarateSpectra}
\input{Figures/B1-dependence}
\section{Experimental Details}

%\MHLnote{I changed the order of the figures since we need a molecular structure earlier.}

The experimental performance of the SLIC, adSLIC, and cSLIC sequences was evaluated by performing singlet-mediated heteronuclear polarization transfer experiments on a solution of [1-$^{13}$C]-fumarate in D$_2$O.
%(see figure~\ref{fig:FumarateSpectra}). 
The heteronuclear pulse sequence scheme shown in Fig.~\ref{fig:FumaratePulseSequences}(a) was used. This has previously been employed to evaluate SLIC and related methods in the context of parahydrogen-enhanced $\mathrm{^{13}C}$ NMR~\cite{eills_singlet_2017}. Although the two \Proton nuclei in [1-$^{13}$C]-fumarate are chemically equivalent, they have different J-couplings with the \Cth nucleus. This J-coupling difference effectively takes the role of the chemical shift difference in spin-1/2 pairs~\cite{eills_singlet_2017,korzeczek_unified_2024}, allowing generation of \Proton singlet order from \Proton transverse magnetisation when a \Proton-SLIC sequence is applied. The J-coupling difference also allows the conversion of \Proton singlet order into \Cth transverse magnetisation when a subsequent SLIC sequence is applied on the \Cth channel. In the absence of dissipation, the optimal duration of the second SLIC pulse is shorter than that of the first SLIC pulse, by a factor of $\sqrt2$ (see ref.~\cite{eills_singlet_2017}).

Experiments were performed on a 200 mM solution of [1-$^{13}$C]-fumarate in D$_2$O at a magnetic field of 9.4~T. The pulse powers were adjusted to provide a nutation frequency of 25.0~kHz on the \Cth channel and 10.0~kHz on the \Proton channel, corresponding to 90\degree pulse durations of 10~$\mu$s and 25~$\mu$s, respectively. 
%For practical reasons, the 
The pulse sequence in figure~\ref{fig:FumaratePulseSequences}(a)  deploys a standard singlet filtration sequence~\cite{tayler_filters_2020,harbor-collins_nmr_2024} in order to suppress any \Cth NMR signals that do not pass through \Proton singlet order between the two SLIC sequences. The experimentally optimised parameters for the sequences were $\{T_{\rm{H}} , T_{\rm{C}}\} = \{524 \text{ ms},370 \text{ ms}\}$ (SLIC), $\{T_{\rm{H}},T_{\rm{C}},n_{\rm{H}},n_{\rm{C}},\alpha\} = \{500 \text{ ms},381 \text{ ms},8,6,0.988\}$  (cSLIC), and $\{T_{\rm{H}}, T_{\rm{C}},\Delta_{\rm{max}}, \xi \} = \{1780 \text{ ms}, 1560 \text{ ms}, 0.5, 0.9\}$ (adSLIC), and $\omega_{J}/(2\pi)= 15$  Hz.%\CBnote{give all details now}.  

\section{Results}
%\MHLnote{(a,b,c,d) is no longer used in Figure~\ref{fig:FumarateSpectra}. Needs changing, after revising the colours.}
Figure~\ref{fig:FumarateSpectra}(orange) shows a \Cth spectrum of [1-$^{13}$C]-fumarate solution generated by a 90\degree \Cth pulse applied to the sample in thermal equilibrium. \Proton decoupling was not applied. The \Cth spectrum shows a triplet multiplet structure due to the J-couplings of the \Cth nucleus to the two fumarate \Proton nuclei, which are nearly magnetically equivalent due to the large $J_\mathrm{HH}$ coupling, the relatively small difference between the two $J_\mathrm{CH}$ couplings, and the very small difference in chemical shifts between the two \Proton sites~\cite{eills_singlet_2017}. 

Figure~\ref{fig:FumarateSpectra}(red) shows a \Cth spectrum obtained by generating \Proton singlet order using a conventional \Proton SLIC sequence of duration $T_\mathrm{H}=524\mathrm{~ms}$, suppressing any other density operator components by a singlet filter sequence, and applying a conventional \Cth SLIC sequence of duration $T_\mathrm{C}=370\mathrm{~ms}$ to generate transverse \Cth magnetisation. The signal intensity is enhanced relative to that in Figure~\ref{fig:FumarateSpectra}(orange), since the \Proton singlet order is generated from thermal-equilibrium \Proton magnetisation, which is approximately four times larger than that of \Cth. The outer multiplet components of the \Cth signal are suppressed, as expected by theory~\cite{eills_singlet_2017}. %\MHLnote{If true, say that the two durations were optimised individually to obtain the maximum signal.}

Figure~\ref{fig:FumarateSpectra}(blue) shows a \Cth spectrum of [1-$^{13}$C]-fumarate obtained under identical conditions, but using adSLIC sequences for both of the SLIC elements in figure~\ref{fig:FumaratePulseSequences}. The adSLIC elements followed the functional form given in Equation \ref{eq:AdPulseShape} and had overall durations of $T_\mathrm{H}=1780\mathrm{~ms}$ and $T_\mathrm{C}=1560\mathrm{~ms}$, with $\Delta_{\rm{max}}=0.5$, $\xi = 0.9$ in both cases. The spectrum shown in Figure~\ref{fig:FumarateSpectra}(blue) was obtained after a sincere attempt to optimise the adSLIC shape parameters on both channels; longer durations led to a deterioration of the performance. 
In this case, only a moderate improvement is evident over the fixed-amplitude SLIC result in Figure~\ref{fig:FumarateSpectra}(red). 

Figure~\ref{fig:FumarateSpectra}(black) shows the \Cth spectrum obtained when using cSLIC for both SLIC elements of the pulse sequence. The parameters used were $T_{H} = 500$ ms, $T_{C} = 381$ ms, $n_{\rm{H}}$ = 8, $n_{\rm{C}}$ = 6, and $\alpha = 0.988$. The spectrum obtained with cSLIC shows a significantly stronger signal than that achieved with both SLIC and adSLIC.

Figure~\ref{fig:B1-dependence} demonstrates the experimental response of the three methods to rf amplitude variations. In all cases the amplitudes of the strong and weak pulses were kept in the same fixed ratio. SLIC displays poor compensation against rf field errors.
Deviations as small as $\pm 10\%$ in the rf amplitude reduce the polarization transfer amplitude by $\sim50\%$. adSLIC displays considerable more robustness with respect rf amplitude deviations, especially on the high-amplitude side. 
%Both SLIC and adSLIC display poor compensation against rf field errors.
%Deviations as small as $\pm 10\%$ in the rf amplitude reduce the polarization transfer amplitude by $\sim50\%$.
In contrast, cSLIC remains largely unaffected by modest rf field variations, retaining half of its peak efficiency even for fractional errors as large as $\pm 50\%$.
The enhanced amplitude of cSLIC over SLIC and adSLIC, as shown in Fig.~\ref{fig:FumarateSpectra}, may thus be attributed to the substantially reduced sensitivity of cSLIC to rf field amplitude variations across the sample volume. %

\section*{Conclusions}

This work demonstrates that cSLIC provides markedly improved compensation against RF field deviations, as compared to %conventional techniques such as 
SLIC and adSLIC, retaining high singlet-order transfer efficiencies even under substantial nutation amplitude mismatches. Although not demonstrated explicitly, this enhanced robustness makes cSLIC particularly attractive for parahydrogen-induced polarisation (PHIP) applications, where maximising heteronuclear polarisation transfer is critical for achieving practical levels of hyperpolarisation. Importantly, cSLIC achieves these advantages as a relatively low-power method: although the central pulse is referred to here as “strong”, it needs to be only modestly more intense than the conventional SLIC element;  typically around five times the SLIC field is sufficient. 
%\MHLnote{good point} 
Counterintuitively, the error-compensation is achieved without increasing the overall sequence duration, in contrast to composite pulse\cite{theis_composite_2014} or optimal control variants of SLIC which are power-efficient but have a lengthened duration~\cite{goodwin_advanced_2017}. The method is also fully compatible with magic-angle (MA) compensation schemes for dipolar field effects, which are important for attaining high molar polarisation in PHIP applications~\cite{dagys_robust_2024,korzeczek_phip_2025}. Additionally, as shown in Section II of the Supporting Information, the performance of the cSLIC sequence can be further enhanced through supercycling, an option that is not readily available for the other methods, and may even be used for DNP. Collectively, these features should make cSLIC a practical tool for enhancing experimental transfer efficiencies in a wide range of NMR experiments exploiting nuclear singlet states and similar phenomena.

\section*{Author contributions}

\textbf{Mohamed Sabba}: Conceptualization (lead), Methodology (lead), Investigation (lead), Data Curation (lead), Software (lead), Visualization (equal), Writing - original draft (equal), Writing - review \& editing (equal \\
\textbf{Christian Bengs}: Visualization (lead), Writing - original draft (lead), Writing - review \& editing (lead) \\
\textbf{Urvashi D. Heramun}: Conceptualization (equal) \\
\textbf{Malcolm H. Levitt}: Conceptualization (equal), Funding acquisition (lead), Supervision (lead), Writing - original draft (equal), Writing - review \& editing (equal)

\section*{Conflicts of interest}
%In accordance with our policy on \href{https://www.rsc.org/journals-books-databases/journal-authors-reviewers/author-responsibilities/#code-of-conduct}{Conflicts of interest} please ensure that a conflicts of interest statement is included in your manuscript here.  Please note that this statement is required for all submitted manuscripts.  If no conflicts exist, please state that ``There are no conflicts to declare''.
MS has received consulting fees from Synex Medical. The other authors have no conflicts to declare.

\section*{Data availability}
The data that support the findings of this study are available from the corresponding author, MS, upon reasonable request.

\section*{Acknowledgements}

We acknowledge funding from the European Research Council (Grant No. 786707-FunMagResBeacons) and EPSRC-UK (Grant Numbers EP/P030491/1, EP/V047663/1, EP/V055593/1, and EP/V047663/1). We acknowledge Matthew S. Rosen and Stephen J. DeVience for sharing valuable insights on the SLIC sequence, and Nino Wili for pointing out similarities with the NOVEL sequence.

\section*{Supporting Information}
The supporting information contains brief details of the theory for multiple use cases of the (c)SLIC sequence, and simulations of supercycled variants of cSLIC.

%%%REFERENCES%%%
\bibliography{References/MS-cSLIC.bib}

\end{document}

\title{Error compensation without a time penalty:  robust  spin-lock-induced crossing in solution NMR}

\author{Mohamed Sabba}
 \email{m.sabba@soton.ac.uk}
\affiliation{Department of Chemistry, University of Southampton, SO17 1BJ, UK}

\author{Christian Bengs}
\affiliation{Department of Chemistry, University of Southampton, SO17 1BJ, UK}

\author{Urvashi Heramun}
\affiliation{Department of Chemistry, University of Southampton, SO17 1BJ, UK}

\author{Malcolm H. Levitt}
 \affiliation{Department of Chemistry, University of Southampton, SO17 1BJ, UK}

\date{\today}

\maketitle

%%%%%%%
%======
%%%%%%%

%================================================
\section{Basic theory of SLIC}

\subsection{Single transition operators}

In order to rationalize SLIC and cSLIC by effective Hamiltonian theory, it is appropriate to define single-transition operators \cite{wokaun_selective_1977,vega_fictitious_1978}. A set of single-transition operators for the subspace of the two arbitrary states $\ket{\alpha}$ and $\ket{\beta}$ could be written as:
\begin{align}
I_{x}^{\alpha,\beta} &= \frac{1}{2} \left(\ket{\alpha}\bra{\beta} + \ket{\beta}\bra{\alpha}\right)  \nonumber 
\\ I_{y}^{\alpha,\beta} &= \frac{1}{2i} \left(\ket{\alpha}\bra{\beta} - \ket{\beta}\bra{\alpha}\right) \nonumber \\
I_{\phi}^{\alpha,\beta} &= \cos\left(\phi \right) I_x^{\alpha, \beta} + \sin\left(\phi \right) I_y^{\alpha,\beta}
 \nonumber \\
I_{z}^{\alpha,\beta} &= \frac{1}{2} \left(\ket{\alpha}\bra{\alpha} - \ket{\beta}\bra{\beta}\right)  \nonumber \\
\mathds{1}^{\alpha,\beta} &=  \ket{\alpha}\bra{\alpha} + \ket{\beta}\bra{\beta}
\end{align}

\subsection{The effective Hamiltonian of SLIC in an AB spin system}

The singlet and triplet states of an isolated spin-1/2 pair are defined as follows:
\begin{align}
\ket{S_0} &= (\ket{\alpha \beta}- \ket{\beta \alpha})/\sqrt{2},
\nonumber\\
\ket{T_+} &= \ket{\alpha\alpha},
\nonumber\\
\ket{T_0} &= (\ket{\alpha \beta}+ \ket{\beta \alpha})/\sqrt{2},
\nonumber\\
\ket{T_-} &= \ket{\beta\beta}
\end{align}

And the Hamiltonian of this spin system, $H_{\rm{AB}}$, can be expressed like so:
\begin{align}
H_{\rm{AB}} \ &= H_\Delta + H_J
\\
H_{\Delta} &= \omega_{\Delta} (I_{1z} - I_{2z} )/2 = \omega_\Delta I_x^{T_0, S_0} 
\\
H_J &= \omega_J I_1 I_2 = \omega_J \left(I_{z}^{T_0, S_0} + I_{1z}I_{2z} \right)
\end{align}

Where $\omega_J = 2\pi J_{12}$ represents the J-coupling and $\omega_{\Delta} = 2\pi \Delta$ represents the chemical shift difference $\Delta = \nu_0^1 - \nu_0^2$.

A key realization is that in the strongly coupled regime ($J_{12} \gg \Delta$), $H_{\Delta}$ should be understood as the \emph{driving interaction}, whereas $H_J$ plays the role of the strong \emph{modulating term} that appears in pulse sequences as a matching condition.

Suppose that the SLIC sequence is defined as follows:
\begin{equation}\label{eq:SLICsequence}
\rm{SLIC_{\pm}}(\tau) = (\pi/2)_{-y} - {[\rm{SL}]}_{\pm x}^{\omega_{\rm{nut}}=\omega_J}(\tau) - (\pi/2)_{+y}
\end{equation}

Where $\rm{[SL]}_{\phi}^{\omega_{\rm{nut}}=\omega_J}(\tau)$ denotes spin-locking of duration $\tau$, phase $\phi$, and nutation frequency $\omega_J$. The final pulse is not necessary at all for the operation of the sequence, but has been inserted to simplify the theory by avoiding working in a tedious tilted frame.

We now evaluate $\tilde{H}_{\Delta}$, the interaction frame Hamiltonian that describes the SLIC sequence, for both positive $(+x)$ and negative $(-x)$ senses of continuous-wave irradiation: 
\begin{align}
\tilde{H}_{\Delta}(\tau) &= \exp\left[+i \frac{\pi}{2} I_{-y} \right] \exp\left[+i \theta_J (I_1 I_2 \pm I_x) \right] H_{\Delta} \exp\left[-i \theta_J (I_1 I_2 \pm I_x) \right]\exp\left[-i \frac{\pi}{2} I_{-y} \right] \nonumber \\
&= \mp \sqrt{2} \pi \Delta \left(I_{x}^{S_0, T_{\pm}} - I_{\omega_J \tau}^{S_0, T_{\mp}} \right) 
\end{align}
And proceed to calculate the effective Hamiltonian $\overline{H}_{\rm{SLIC}_{\pm}}^{(1)}$ over a cycle corresponding to a full J-period $T = 1/J$:
\begin{align}
\overline{H}_{\rm{SLIC}}^{(1)} &= \frac{1}{T} \int_{0 }^{T } \tilde{H}_{\Delta}(\tau) \,d\tau,~T=1/J \nonumber \\
&= \mp \sqrt{2} \pi \Delta I_{x}^{S_0, T_\pm }
\end{align}

The most critical point here is that the mechanism of the SLIC sequence can be fully explained by the above effective Hamiltonian, which engineers a transition between the state $\ket{S_0}$ and either of the triplet states $\ket{T_{\pm}}$, depending on the phase of the SLIC pulse. We now proceed to "translating" this picture to other applications of SLIC.

\subsection{Other contexts where SLIC (and cSLIC) are applicable}

Rather than derive multiple effective Hamiltonians for the various different situations in which SLIC (and analogous sequences such as NOVEL) are applicable on a case-by-case basis, we find it more convenient to provide a table containing a list of relevant transitions $\ket{r} \leftrightarrow \ket{s}$, rf amplitude matching conditions (or resonances) in terms of the modulation frequency $\Omega_\mu$, and driving terms characterized by a Rabi frequency $\omega_u$. \\ \\

\begin{table}[tbh!]
\begin{tabular}{c|p{30 mm}|p{30 mm}|p{40 mm}|p{40 mm}}
Sequence     &  Transition & Matched rf \newline amplitude ($\omega_{\rm{rf}}$) & Driving term \newline [Rabi frequency, $\omega_\mu$] & Modulating term \newline [Resonant frequency, $\Omega_\mu$] \\ \hline
SLIC (AB) & $\ket{S_0} \leftrightarrow \ket{T_{\pm}}$ & $\omega_{\rm nut}^{12}$ & chemical shift difference \newline $[\sqrt{2}\pi\Delta]$  & J-coupling \newline $ [2\pi J_{12}]$ \\ \hline
SLIC (PHIP) & $\ket{\alpha^{\rm{S}} S_0^{\rm{I}}} \leftrightarrow \ket{\beta^{\rm{S}} T_{\pm}^{\rm{I}}}$ & $\omega_{\rm nut}^S $ & differential coupling \newline $[\pi(J_{IS}- J_{I'S})/2]$ & homonuclear J-coupling \newline $[2\pi J_{II} ]$\\ \hline 
SLIC (AA'XX') & $\ket{T_{\pm}^{A}S_0^{X}} \leftrightarrow \ket{S_{0}^{A}T_0^{X}} $ & $\omega_{\rm nut}^A$ & out-of-pair \newline coupling difference \newline $[\pi(J_{AX}- J_{AX'})/\sqrt{2}]$ & in-pair \newline coupling (difference) \newline $[2\pi(J_{AA} - J_{XX'})]$
\\ \hline
SLIC (AA'XX') & $\ket{T_{\pm}^{A}T_0^{X}} \leftrightarrow \ket{S_{0}^{A}S_0^{X}} $ & $\omega_{\rm nut}^A $ & out-of-pair \newline coupling difference \newline $[\pi(J_{AX}- J_{AX'})/\sqrt{2}]$ & in-pair \newline coupling (sum) \newline $[2\pi(J_{AA} + J_{XX'})]$
\\ \hline
NOVEL & $\ket{\alpha^{\rm{e}} \beta^{\rm{n}}} \leftrightarrow \ket{\beta^{\rm{e}} \alpha^{\rm{n}}}$ \newline or \newline $\ket{\alpha^{\rm{e}} \alpha^{\rm{n}}} \leftrightarrow \ket{\beta^{\rm{e}} \beta^{\rm{n}}}$& $\omega_{\rm{nut}}^{\rm{e}}$ & elecron-nuclear coupling \newline [$\pi J_{\rm{e n}}/2$] &  nuclear Larmor frequency \newline [$\omega_{0}^{\rm{n}}$]\\
\end{tabular}
\caption{A table of various analogues of the SLIC pulse sequence. All sequences are fully characterized by their particular target transitions, the matching condition $\omega_{\rm{rf}} = \Omega_{\mu}$ relating a nutation frequency to the modulation frequency $\Omega_{\mu}$, and driving terms which allow population transfer with a Rabi frequency $\omega_{\mu}$.}
\label{tab:Sequences}

\end{table}

\subsection{Generalized response of SLIC to rf amplitude errors}

In the most general case, the dependence of the excitation efficiency of SLIC on rf amplitude errors $\epsilon_{\rm{rf}}$ corresponds to the familiar textbook case of a detuned Rabi oscillation:

\begin{equation}\label{eq:SLICresponse}
\xi_{\rm{SLIC}}(t) =\sin ^2(\theta_\mu ) \sin ^2\left(\frac{1}{2} \omega_\mu \csc (\theta _\mu)t \right)
\end{equation}

Where the "detuning angle" $\theta_\mu$ is defined in terms of the Rabi frequency $\omega_\mu$ and the effective detuning frequency $\Omega_{\mu}$: 

\begin{align}
\theta_{\mu} = \arctan\left(\frac{\omega_{\mu}}{\Omega_{\mu} \epsilon_{\rm{rf}}}\right) \nonumber \\
\end{align}

For a nominal duration $t = \pi/\omega_\mu$ the efficiency is simply:

\begin{equation}\label{eq:SLICresponsenom}
\xi_{\rm{SLIC}}(\pi/\omega_\mu) = \frac{\pi^2}{4} \rm{sinc} ^2 \left(\frac{\pi}{2} \csc (\theta _\mu)\right)
\end{equation}

Which is a narrowband sinc-squared response, depicted in Figure \ref{fig:rferrordeps}. The extreme sensitivity of the SLIC sequence (and analogues such as NOVEL) to the matching condition presents a major source of inconvenience to the variety of experiments mentioned beforehand.

\subsection{The effective Hamiltonian of the cSLIC sequence in an AB spin system}

The effective Hamiltonian of the cSLIC sequence, as shown in the appendix of another reference \cite{heramun_spinor_2026}, is given by the following in the limit of a negligibly short compensating pulse ($\alpha \rightarrow 1$): 
\begin{align}
\overline{H}^{(1)}_{\rm{cSLIC}} &= \lim_{\alpha \rightarrow 1} \int_0^{1/J} \tilde{H}_{\Delta}(t) ~\rm{d}t \nonumber\\
&= 
-\sqrt{2}\pi\Delta
\big[
\mathrm{sinc}\left(f_{+}\right) I_x^{S_0, T_+} - \mathrm{sinc}\left(f_{-}\right) I_x^{S_0, T_{-}}
\big]
\end{align} 
where the arguments of the $\mathrm{sinc}$ functions are:
\begin{align}
f_{+} &= \pi \epsilon_{\rm{rf}} \nonumber \\
f_{-} &= \pi \left(2+\epsilon_{\rm{rf}}\right)
\end{align}
These represent a pair of counter-rotating frequency components, centred at the resonances of the two nominal SLIC matching conditions: $\omega_{\rm{nut}}^{\rm{SLIC}} = +\omega_J$ ($\epsilon_{\rm{rf}} =0$), and $\omega_{\rm{nut}}^{\rm{SLIC}} = -\omega_J$ ($\epsilon_{\rm{rf}} =-2$).

\subsection{Generalized response of cSLIC as a function of rf amplitude errors}

In the most general case, the excitation efficiency of cSLIC is given by the following function:

\begin{equation}
\xi_{\rm{cSLIC}}(t)= \frac{f_-^2 - f_+^2}{f_-^2+f_+^2}\sin ^2\left(\frac{1}{2} \omega_\mu    \sqrt{\rm{sinc}^2\left(f_-\right)+\rm{sinc}^2\left(f_+\right)} t\right)
\end{equation}

When compared to Equation \ref{eq:SLICresponse}, the dependence on the $\Omega_\mu$ term has vanished. The dependence on $\epsilon_{\rm{rf}}$ is solely due to interference between the two matching conditions $f_-$ and $f_+$. 

For a nominal duration $t=\pi/\omega_{\mu}$, the excitation efficiency of cSLIC becomes:

\begin{equation}\label{eq:cSLICresponsenom}
\xi_{\rm{cSLIC}}(t=\pi/\omega_{\rm{\mu}}) =\frac{f_-^2 - f_+^2}{f_-^2+f_+^2}\sin ^2\left(\frac{\pi}{2}    \sqrt{\rm{sinc}^2\left(f_-\right)+\rm{sinc}^2\left(f_+\right)} \right)
\end{equation}

\begin{figure}
    \centering
    \includegraphics[width=1.0\linewidth]{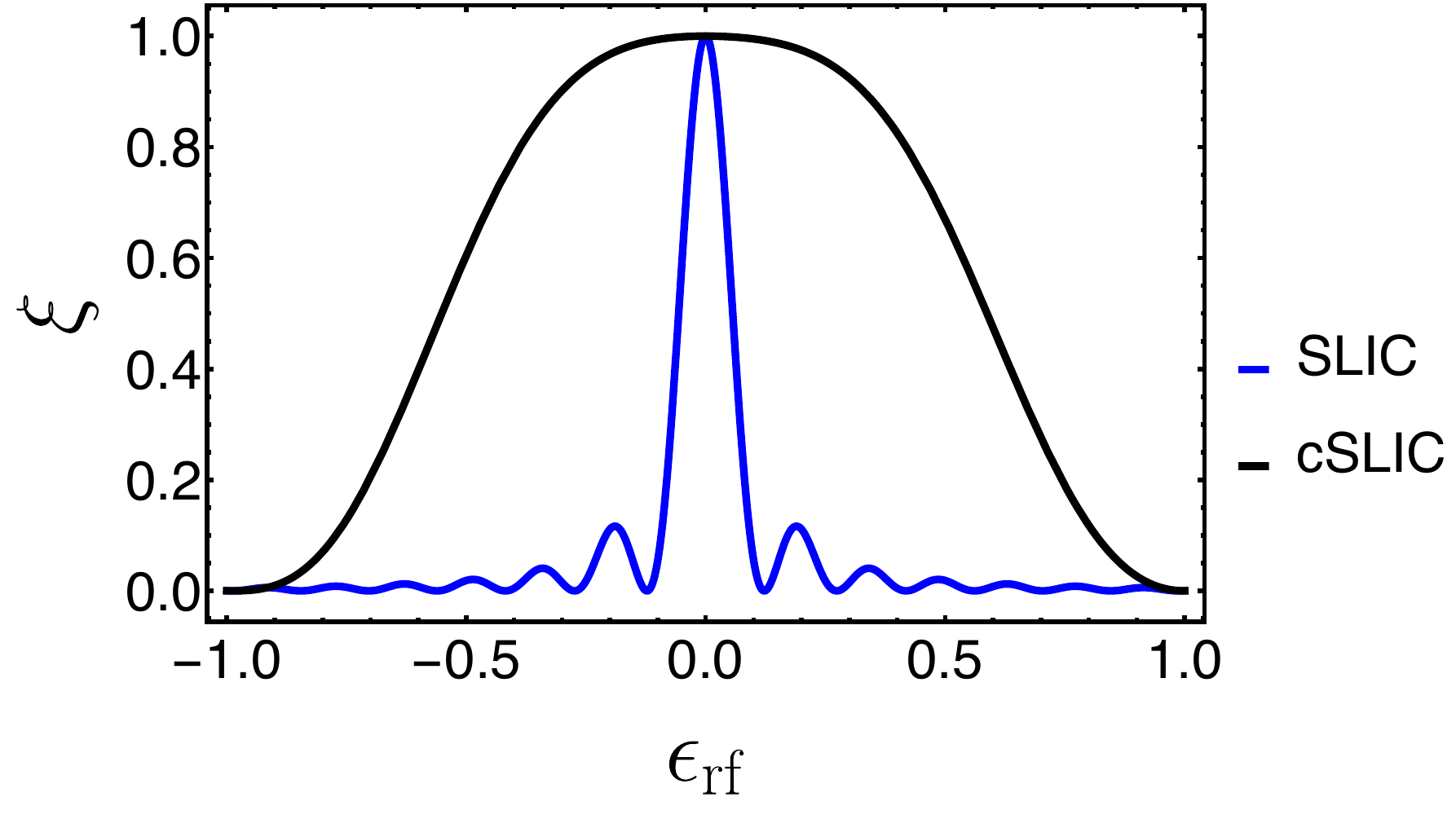}
    \caption{Analytical excitation efficiences $\xi$ of the SLIC and cSLIC sequences as a function of rf amplitude error $\epsilon_{\rm{rf}}$, using equations \ref{eq:SLICresponsenom} and \ref{eq:cSLICresponsenom}. The parameters used for SLIC are the Rabi frequency $\omega_{\mu} = \omega_{\Delta}/\sqrt{2}$ and $\Omega_\mu = \omega_J$, with $\omega_{\Delta} = 2\pi \times10$ and $\omega_J = 2\pi\times100$.}
    \label{fig:rferrordeps}
\end{figure}

A comparison of the analytical equations describing the rf error dependence of SLIC (equation \ref{eq:SLICresponsenom}) and cSLIC (equation \ref{eq:cSLICresponsenom}) is shown in Figure \ref{fig:rferrordeps}. The cSLIC sequence (or what would be the analogous "cNOVEL" sequence in the context of DNP) has a broadband response whereas the SLIC sequence has a narrowband response that worsens for larger values of $\Omega_\mu / \omega_\mu$.

\section{Supercycled variants of \MakeLowercase{c}SLIC}

$\mathcal{C}$, the cyclic element building block of the cSLIC pulse sequence used in this paper, can be expressed in terms of the two inversion elements $\rm{A}$ and $\rm{B}$:
\begin{align}
\text{A} &= 180_x^{\omega_{\rm{nut}}=\omega_J} \nonumber \\
\text{B} &= 180_{-x}^{\omega_{\rm{nut}}=\omega_{\rm{strong}}} 
\end{align}
The position of the compensating pulse element "BB" is not particularly important; it may be placed at the beginning, middle, or end of the pulse sequence without affecting the performance, and indeed all of the following cyclic elements may be used to compose the cSLIC sequence:
\begin{align}
\mathcal{C}_1 &= \text{AABB} \\
\mathcal{C}_{2} &= \text{ABBA} \\
\mathcal{C}_{3} &= \text{BBAA}
\end{align}
The variant of cSLIC used in this paper (ABBA) was chosen purely for aesthetic reasons, and as a tribute to the Swedish pop band that bears the same moniker.

Combining the different permutations $C_i$ can be used to improve the performance of cSLIC as a function of resonance offset, at the expense of increased sensitivity to rf amplitude errors. For example, some "supercycles" \cite{burum_lowpower_1981,levitt_symmetry_2008} with better off-resonance performance (denoted $S_i$) are:
\begin{align}
S_1 &= \mathcal{C}_2 \nonumber \\
S_2 &= \mathcal{C}_1\mathcal{C}_2 \nonumber \\
S_3 &= \mathcal{C}_1\mathcal{C}_2\mathcal{C}_3
\end{align}

\begin{figure}
    \centering
    \includegraphics[width=1\linewidth]{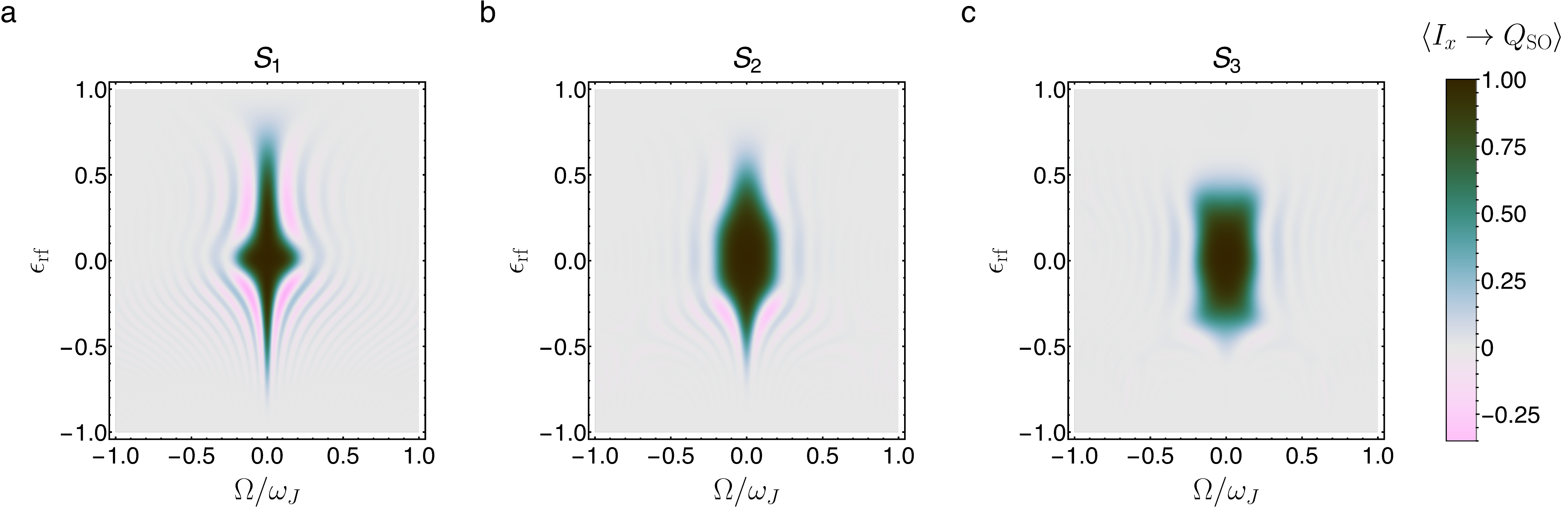}
    \caption{Numerical simulations showing the performance of supercycled variants of cSLIC as a function of rf amplitude errors $\epsilon_{\rm{rf}}$ and relative resonance offset errors $\Omega/\omega_{J}$. Panel (a): the primitive cycle $S_1 =$ ABBA. Panel (b): the 8-step supercycle $S_2=$ AABBABBA. Panel (c): the 12-step supercycle $S_3 =$ AABBABBABBAA. The simulations are performed for a 2-spin-1/2 system with $\omega_J = 2\pi\times100$, and $\omega_{\Delta} = 2\pi \times3$, assuming $\alpha = 0.99$.}
    \label{fig:supercycles}
\end{figure}

The performance of these supercycled variants is shown in Figure \ref{fig:supercycles}. The $S_3$ variant of cSLIC shows more uniform performance, with no generation of negative singlet order, at the expense of a smaller on-resonance $\epsilon_{\rm{rf}}$ bandwidth.

\clearpage

\
\
\bibliography{References/SI}%

%% file: Symbols/main.tex
\input{Symbols/MHL-symbols}
\input{Symbols/MS-symbols}

%% file: Symbols/MHL-symbols.tex
\newcommand{\blue}[1]{\textcolor{blue}{#1}}
\newcommand{\red}[1]{\textcolor{red}{#1}}
\newcommand{\MHLToHere}{\blue{\ \newline ***MHL TO HERE ***\ \newline}}
\newcommand{\MHLnote}[1]{\blue{[MHL: #1]}}
\newcommand{\orange}[1]{\textcolor{orange}{#1}}
\newcommand{\CBnote}[1]{\orange{[CB: #1]}}
%=================
\newcommand{\Proton}{$\mathrm{^1H}$\xspace}
\newcommand{\Cth}{$\mathrm{^{13}C}$\xspace}
\newcommand{\degree}{$^\circ$\xspace}

%% file: Symbols/MS-symbols.tex
%Mohamed symbols
\newcommand{\cyan}[1]{\textcolor{cyan}{#1}}
\newcommand{\SBnote}[1]{\cyan{[MS: #1]}}
%===
\newcommand{\Rseq}[3]{{#1}_{#2}^{#3}}
%===

%% file: Figures/ContourPlots.tex
\begin{figure*}[bt]
\centering
\includegraphics[width=\linewidth,trim={0 0cm 0 0},clip]{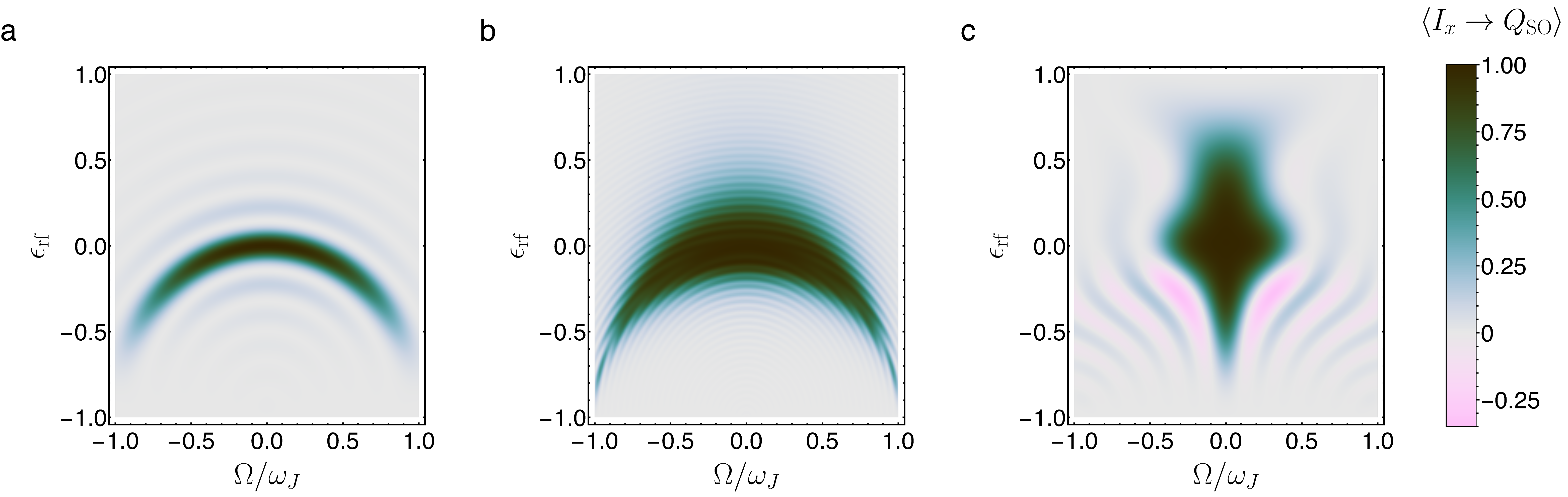}
\caption{Contour plots for the transformation amplitude of transverse magnetisation into singlet order, defined by $\langle I_{x}\rightarrow Q_{\rm SO}\rangle={\rm Tr}\{Q_{\rm SO}U I_{x}U^{\dagger}\}/{\rm Tr}\{Q_{\rm SO}Q_{\rm SO}\}$, where $U$ represents the propagator for a specific SLIC element. Results are shown for (a) SLIC, (b) adSLIC, (c) cSLIC against resonance offset (horizontal axis) and deviations in the rf amplitude (vertical axis).
%\MHLnote{In the text I have assumed this order, which is different to our discussion.}
%The plotted quantity $\vert a_{\rm{M\rightarrow S}}\vert$ is the simulated transformation amplitude of transverse magnetisation $I_x$ into nuclear singlet order $-\tfrac{4}{3}\mathbf{I}_j\cdot\mathbf{I}$.
All simulations are performed for a two-spin system with $J = 15$ Hz and $\Delta = 1.9$ Hz. The total durations of the elements $T$ are: 375 ms (SLIC and cSLIC) and 1560 ms (adSLIC) respectively. For cSLIC, the repetition number is $n=6$ and $\alpha=0.99$. For adSLIC, the rf-amplitude modulation is given by Equation \ref{eq:AdPulseShape} with $\Delta_{\rm{max}} = 0.5$ and $\xi = 0.9$.
}
\label{fig:ContourPlots}
\end{figure*}

%% file: Figures/cSLIC.tex
\begin{figure}[tb]
\centering
\includegraphics[width=\columnwidth,trim={0 7cm 15.5cm 0},clip]{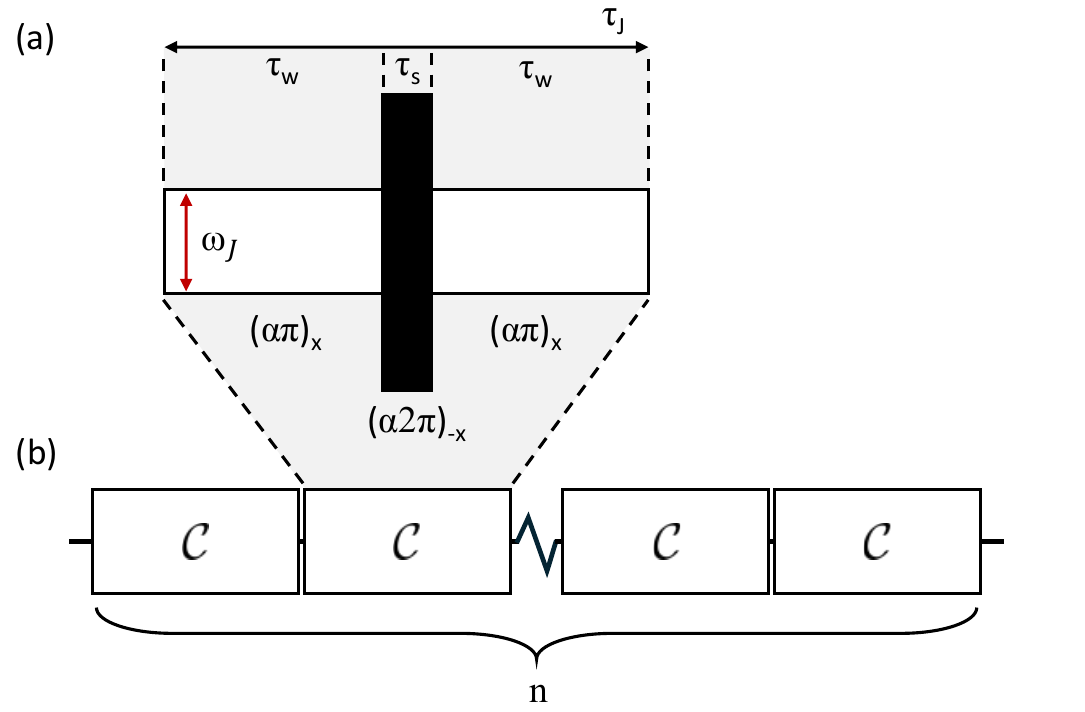}
%trim={0 7cm 15.5cm 0},clip
\caption{Pulse sequence schematic for cSLIC. (a) The basic element consists of a concatenation of two weak pulses separated by a strong pulse. The weak pulses are of amplitude $\omega^{\rm weak}_{\rm nut}=\omega_{J}$, duration $\tau_{w}$ and produce a net rotation of $(\alpha \pi)$ along the $x$-axis. The strong central pulse is of amplitude $\omega^{\rm strong}_{\rm nut}>\omega_{J}$, duration $\tau_{s}$ and produces a net rotation of $(\alpha 2\pi)$ along the $-x$-axis. The parameter $\tfrac12\leq\alpha\lesssim1$ is defined by equation~\ref{eq:alpha factor}. 
The pulse durations are constrained by $2\tau_{w}+\tau_{s}=\tau_{J}$. (b) cSLIC-based singlet excitation consists of $n=\lfloor J/(\sqrt2\Delta)\rceil$ repetitions of the basic element shown in (a). 
}
\label{fig:cSLIC}
\end{figure}

%% file: Figures/FumaratePulseSequences.tex
\begin{figure}[bt]
\centering
\includegraphics[width=\columnwidth,trim={0 2.5cm 7.5cm 0},clip]{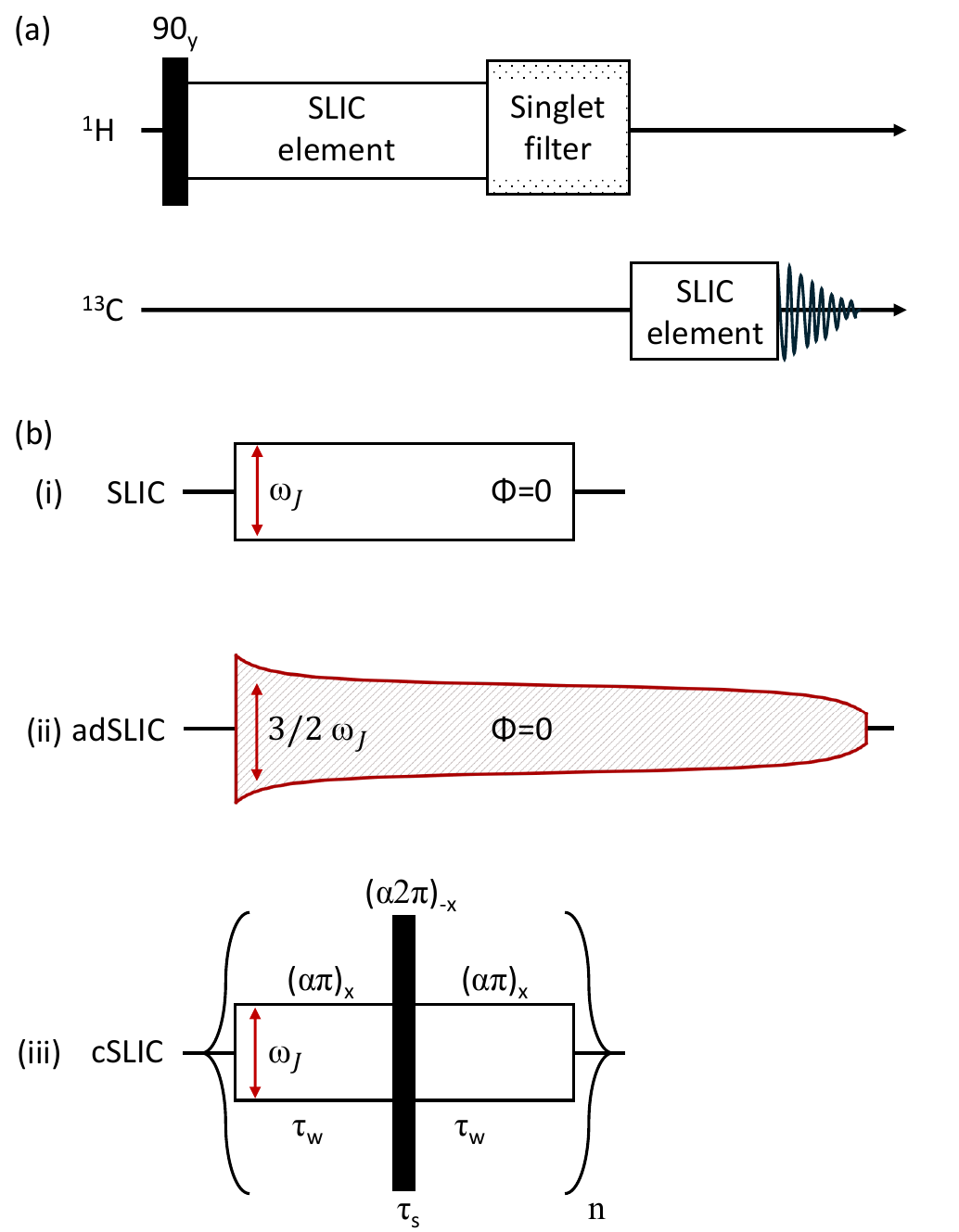}
\caption{(a) Pulse sequence for heteronuclear polarization transfer from \Proton to \Cth, through an intermediate \Proton singlet state, in systems of two \Proton nuclei and one \Cth nucleus, as in figure~\ref{fig:FumarateSpectra}.
An initial $90^{\circ}_{y}$ pulse generates transverse proton magnetisation, followed by a singlet preparation element using one of the SLIC variants shown in (b). After singlet-order preparation, a filter element removes any spurious density-operator terms. A second SLIC variant transforms singlet order into heteronuclear magnetisation. %The same type of SLIC element was always used in both halves of the pulse sequence. The duration of the proton singlet element is approximately $\sim$ 1.4 times as large as the duration of the carbon singlet element, $T_{\rm H}\simeq \sqrt{2}T_{\rm C}$ (see ref.~\cite{eills_singlet_2017} for details). 
(b) SLIC elements used in this work. (i) The basic SLIC element consists of a single $x$ pulse with amplitude $\omega_{J}$. (ii) Adiabatic SLIC consists of an amplitude-modulated $x$ pulse following the functional form given in Equation \ref{eq:AdPulseShape}.
(iii) The cSLIC element follows the procedure outlined in fig.~\ref{fig:cSLIC}. 
%\MHLnote{plz make the depicted shape exactly the same as that used experimentally.}
} 
\label{fig:FumaratePulseSequences}
\end{figure}

%% file: Figures/FumarateSpectra.tex
\begin{figure}[tb]
\centering
\includegraphics[width=\linewidth,trim={0 8.96cm 17.7cm 0},clip]{Graphics/FumarateSpectra.pdf}
\caption{\Cth spectra of [1-$^{13}$C]-fumarate dissolved in $\mathrm{D_2 O}$. Inset shows the chemical structure and the J-coupling parameters.
Fourier transform (green) of the free-induction decay generated by a single 90\degree \Cth pulse applied to a sample in thermal equilibrium. \Cth spectra for SLIC (pink), adSLIC (blue), and cSLIC (black) were obtained using the pulse sequence strategy shown in Figure~\ref{fig:FumaratePulseSequences}. 0.5 Hz of line broadening was applied to all spectra. The pulse sequence parameters are summarised in the experimental details.
}
\label{fig:FumarateSpectra}
\end{figure}

%% file: Figures/B1-dependence.tex
\begin{figure}
    \centering \includegraphics[width=\linewidth,trim={0 7.55cm 15.65cm 0},clip]{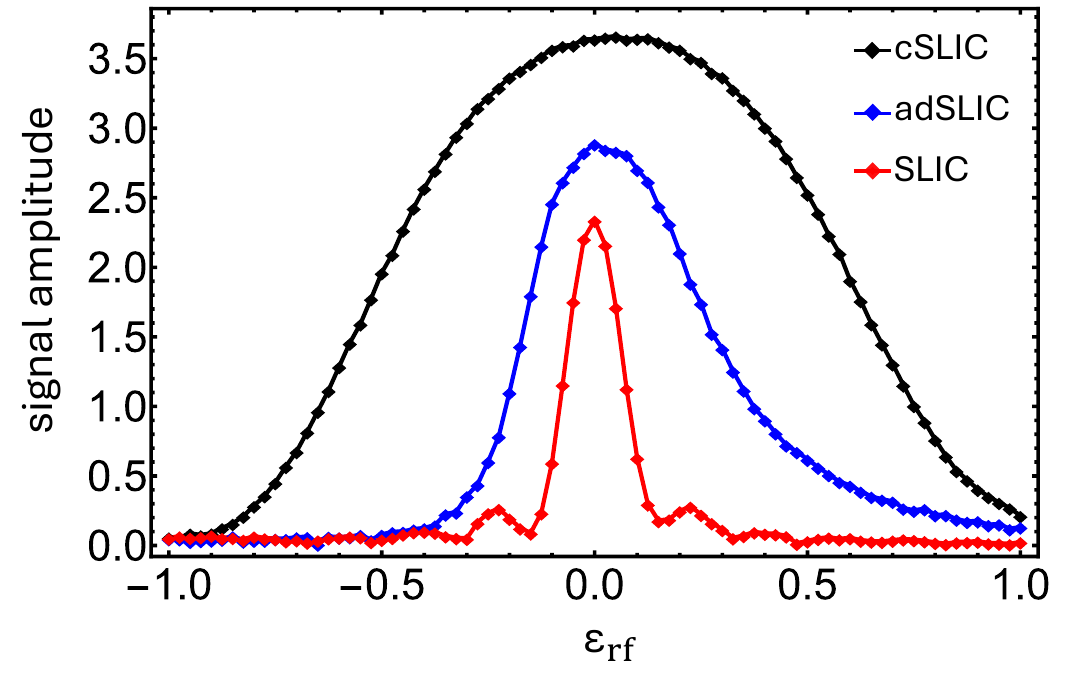}
    \caption{Experimental \Cth signal amplitude
    for singlet-mediated heteronuclear polarisation transfer in [1-$^{13}$C]-fumarate, 
     as a function of the fractional \Cth nutation amplitude mismatch, defined in Equation~\ref{eq:FractionalrfDeviation}. 
    Signal amplitudes have been normalised against the amplitude of the central peak in the $90$\degree spectrum.} 
\label{fig:B1-dependence}
\end{figure}